\documentclass[preprint,aps,showpacs]{revtex4}
\usepackage{graphicx,amsmath}

\usepackage[english]{babel}

\begin{document}

\title{Microwave Response of V$_3$Si Single Crystals:
Evidence for Two-Gap Superconductivity}

\author{Yu.A.~Nefyodov, A.M.~Shuvaev, and M.R.~Trunin}
\affiliation{Institute of Solid State Physics RAS, 142432
Chernogolovka, Moscow district, Russia}

\begin{abstract}
The investigation of the temperature dependences of microwave
surface impedance and complex conductivity of V$_3$Si single
crystals with different stoichiometry allowed to observe a number
of peculiarities which are in remarkable contradiction with
single-gap Bardeen-Cooper-Schrieffer theory. At the same time,
they can be well described by two-band model of superconductivity,
thus strongly evidencing the existence of two distinct energy gaps
with zero-temperature values $\Delta_1\approx 1.8\,T_c$ and
$\Delta_2\approx 0.95\,T_c$ in V$_3$Si.
\end{abstract}
 \maketitle

The story of multi-gap superconductors goes to the middle of the
last century, when the extension of Bardeen-Cooper-Schrieffer
(BCS) theory~\cite{BCS} was proposed~\cite{Mosk,Suhl}. The
followed experimental investigations, however, contradicted each
other showing the existence of single gap, multiple gaps and
slightly anisotropic gap in various superconducting materials. The
interest to this phenomenon has been stimulated by the discovery
of two-gap superconductivity in MgB$_2$. The existence of at least
two different energy bands crossing the Fermi-level (particular
feature of MgB$_2$) appears to be the prerequisite for the
observation of multiple gaps. The second requirement, as follows
from~\cite{Suhl}, is a weak interband scattering. Such processes
can be significantly reduced if wave functions of electrons from
two bands have different symmetry. For example this may happen
when energy band structure has both flat and non-flat areas near
the Fermi level. Flat areas lead to a singularity in the density
of states at the Fermi level and can be experimentally detected
for instance by non-linear temperature dependence of the
resistivity $\rho(T)$ in the normal state. In the opposite case of
more or less similar bands structures even small amount of
impurities leads to high interband scattering and almost excludes
the possibility to detect multi-gap response of the material.

Apparently, the above requirements apply to the layered
superconductor NbSe$_2$ and to A15 structure superconductors
Nb$_3$Sn, V$_3$Si, V$_3$Ga, in which the density of states has
very high and narrow peak just in the vicinity of the Fermi level
according to band-structure calculations~\cite{Papa}. Recently the
existence of multiple gaps in Nb$_3$Sn polycrystalline sample has
been proposed to explain specific heat
measurements~\cite{Bouquet}. The authors of~\cite{Boak} showed the
similarity of the magnetic field dependence of thermal
conductivity in NbSe$_2$ to that of MgB$_2$ and concluded about
the presence of the second energy gap in NbSe$_2$. Such a
similarity was not found in V$_3$Si. However, back in 1969 J.Brock
denoted the existence of the second gap as one of the possible
explanations of the peculiarities seen in the specific heat of
V$_3$Si~\cite{Brock}.

The measurements of the temperature dependences of microwave
conductivity both in low-$T_c$ and high-$T_c$ superconductors were
very informative. They proved the applicability of BCS theory to
conventional superconductors, allowed to distinguish
superconductors with different order parameter symmetry and to
measure the values of energy gap, penetration depth,
quasiparticles relaxation rate, mean free path, that are the
quantities important for comparison with first-principles
calculations. In the $c$-axis oriented MgB$_2$ films an anomalous
peak in the real part of conductivity around $T/T_c=0.6$ was
observed~\cite{Jin}. Its origin is associated with the smallness
of the gap in a dirty two gap superconductor. However, up to now
there was no successful preparation of MgB$_2$ samples series with
different impurity concentration and falling into dirty and pure
limit. A15 materials which preparation technology is well
established seem to be the best candidates for the detailed study
of two-gap superconductivity.

In this Letter we present investigation of the temperature
dependences of microwave surface impedance $Z(T)=R(T)+iX(T)$ and
complex conductivity $\sigma(T)=\sigma'(T)-i\sigma''(T)$ of
V$_3$Si crystals with different degree of imperfection and
critical temperatures. The activation energy $\Delta_2$ of
low-temperature dependences $\delta X\propto\exp(-\Delta_2/T)$
amounts to $\Delta_2(0)\approx 0.95\,T_c$, that is much smaller
than BCS value $\Delta(0)=1.76\,T_c$. The so-called coherent peak
appearing in $\sigma'(T)$ of less perfect sample is located at
$T\approx T_c/2$, that is lower than BCS value $T\approx
0.85\,T_c$. The curve of $\sigma'(T)$ in the clean sample has no
peak at all. The dependences $\sigma''(T)$ of both samples
demonstrate an evident inflection at temperature $T\approx T_c/2$.
The observed peculiarities are in remarkable contradiction with
single-band BCS model but can be well fitted by two-band model of
superconductivity thus evidencing the existence of two distinct
energy gaps with zero-temperature values $\Delta_1\approx
1.8\,T_c$ and $\Delta_2\approx 0.95\,T_c$ in V$_3$Si.

Two single crystals of V$_3$Si were investigated. Sample \#1 had
stoichiometric composition, that is 25\% of silicon, whereas \#2
was prepared with 1\% of silicon deficiency. Both samples having
rectangular shape were preliminary polished and cleaned to remove
possible surface contamination. Measurements of the
$ac$-susceptibility at 100~kHz showed the superconducting
transition temperatures $T_c=16.5$~K and $T_c=12.5$~K for samples
\#1 and \#2, respectively, and the transitions widths less than
0.2~K.

The microwave experiments were performed by the "hot-finger"
technique~\cite{Sri7,Tru1}. We placed the sample into the center
of the cylindrical superconducting niobium cavity resonating at
the frequency $f=\omega/2\pi=9.4$~GHz in TE$_{011}$ mode and
having high unloaded quality factor $Q_0\simeq 2\times 10^7$.
Since the sample is located at the antinode of the microwave
magnetic field, the resonance frequency $f(T)$ of the system and
its quality factor $Q(T)$ can be easily related to the surface
impedance $Z(T)=R(T)+iX(T)$ of the sample using simple
formulas~\cite{Tru1}. The real part, surface resistance $R(T)$, is
proportional to the microwave power absorbed by the sample,
whereas the imaginary part, surface reactance $X(T)$, is the
reactive component which defines the electromagnetic field
screening. In the superconducting state when $R(T)\ll X(T)$
surface reactance directly gives the magnetic field penetration
depth $\lambda(T)=X(T)/\omega\mu_0$, where $\mu_0=4\pi\times
10^{-7}$~H/m. The main three advantages of this technique are (i)
the sensitivity in the superconducting state of even small crystal
being high enough to detect the change in $\lambda(T)$ amounting
to parts of a nanometer, (ii) the possibility to obtain the value
of $\lambda(0)$ and to measure (iii) the normal state
conductivity.

The results of the surface impedance measurements in the
temperature range $2\le T\le 100$~K are shown in Fig.~1. In the
normal state the data on both samples demonstrate the equality
$R(T)=X(T)$ from $T\simeq 35$~K and up to 300~K thus conforming
the validity of the normal skin effect. This makes possible to
obtain the resistivities $\rho(T)=2R^2(T)/\omega\mu_0$ shown in
the inset to Fig.~1. Nonlinear temperature dependence of $\rho(T)$
in the entire temperature range points to a singularity in the
density of states in V$_3$Si. Low value of the resistivity
$\rho(T_c)\simeq 2$~$\mu\Omega\cdot$cm is indicative of high
purity of the sample \#1. The non-stoichiometric sample \#2 shows
20 times higher resistivity $\rho(T_c)$.

A quasiparticle relaxation time $\tau$ near the transition
temperature can be estimated from the standard formula of the
Gorter and Casimir two-fluid model
$\omega\tau(T_c)=X^2(0)/2R^2(T_c)$: $\omega\tau(T_c)\simeq 0.06$
and 0.006 for samples \#1 and \#2, respectively. Rather large
value of $\tau$ in the sample \#1 leads to the non-zero value of
the imaginary part of the normal state conductivity (time
dispersion of the Drude conductivity) and causes the discrepancy
between $R(T)$ and $X(T)$ curves in Fig.~1 at $17<T<35$~K. We get
parameter $\gamma=1/\tau(T_c)$: $\gamma\simeq 7$~K for sample \#1
and $\gamma\simeq 70$~K for sample \#2. From $T_c$ values we can
estimate the amplitudes of the superconducting gaps using BCS
relation $\Delta(0)=1.76\,T_c$. We have $\Delta(0)\simeq 30$~K for
sample \#1 and $\Delta(0)\simeq 20$~K for sample \#2. Combining
the above obtained values we conclude that sample \#1 can be
classified as clean superconductor ($\Delta>\gamma$), whereas
sample \#2 falls into the dirty limit ($\Delta<\gamma$).

In Fig.~2 the dependences of $\ln[X(T)-X(0)]$ on $T_c/T$ are
shown. Since in the London superconductor [$X(T)-X(0)]\propto
\exp[-\Delta(0)/T]$, the slopes of the linear at $T<T_c/3$
sections of these graphs (solid lines in Fig.~2) directly give us
the values $\Delta_2(0)/T_c=0.97$ and $\Delta_2(0)/T_c=0.93$ for
samples \#1 and \#2, respectively. These values are approximately
half in comparison with the standard BCS value.

In the superconducting state we find the values of
zero-temperature penetration depth $\lambda(0)=X(0)/\omega\mu_0$
being equal 150~nm and 180~nm for samples \#1 and \#2,
respectively. V$_3$Si can be classified as London superconductor
in which the coherence length $\xi(0)\approx 5$~nm~\cite{Boak} at
$T=0$ is much less than the magnetic field penetration depth. So,
the relationship between the impedance and complex conductivity is
given by local formula $\sigma=i\omega\mu_0/(R+iX)^2$. To obtain
the temperature dependences of complex conductivity we use the
values $X=X(T)$ and $R=R(T)-R_{res}$, where $R(T)$ and $X(T)$ are
shown in Fig.~1 and $R_{res}=R(T\to 0)$ is so-called residual
surface resistance~\cite{Halbr}. Below 4~K surface resistance
curves of both samples rich the plateau which enables us to
determine $R_{res}$ values amounting to $1.7\pm 0.02$~m$\Omega$
and $0.24\pm 0.02$~m$\Omega$ in samples \#1 and \#2, respectively.
The account for $R_{res}$ does not change $\sigma''(T)$ but
influences $\sigma'(T)$.

The inflection seen in $X(T)$ curve in Fig.~1 at $T\approx 10$~K
for the sample \#1 becomes more evident when looking at the
imaginary part $\sigma''(T)$ of the conductivity shown in Fig.~3.
Moreover, quite similar inflection persists also in the second
sample. No such an inflection is predicted by the single-gap BCS
model (dashed and dotted lines in Fig.~3). The accuracy of our
method (less than 1\,nm for $\delta\lambda(T)$) and the fact that
we directly extract the absolute value $\lambda(0)$ of the
penetration depth suggest that the peculiarities of $\sigma''(T)$
curves can not be caused by an experimental error. For the sake of
comparison in the same figure we show also our measurements of the
conductivity $\sigma''(T)$ of Pb$_{2-x}$Mo$_6$S$_8$ crystal
(Chevrel phase) which is well fitted by BCS theory.

Fig.~4 demonstrates the temperature dependences of the real part
of the conductivity $\sigma'(T)$. It is well known that in the
dirty limit of BCS model there is a pronounced coherence peak in
$\sigma'(T)$ dependence. Its maximum corresponds to the condition
$\Delta(T)\approx T$ which happens at $T\approx 0.85\,T_c$. The
dotted line in Fig.~4 shows $\sigma'(T)$ curve for dirty BCS
superconductor with $T_c=12.5$~K. However, dirty sample \#2 has a
broad peak centered at $T\approx 0.5\,T_c$. Its position is only
slightly affected by the choice of $R_{res}$ and by no means can
be shifted to the BCS value. The curve of $\sigma'(T)$ in sample
\#1 has no peak because the coherence peak disappears in the clean
limit of the BCS model (dashed line in Fig.~4).

The peculiarities of $\sigma(T)$ curves in Figs.~3 and 4 can be
explained in the framework of simplified two-band model with free
parameters of intraband ($g_{11}$ and $g_{22}$) and interband
coupling ($g_{12}$ and $g_{21}$, chosen to be equal for
simplicity). In the model~\cite{Mosk,Suhl} the values of
superconducting gaps $\Delta_i$ in two bands can be obtained from
the following set of equations:
\begin{eqnarray} \Delta_1=g_{11}\Delta_1
I(\Delta_1)+g_{12}\Delta_2 I(\Delta_2),&&\qquad
\Delta_2=g_{21}\Delta_1 I(\Delta_1)+g_{22}\Delta_2 I(\Delta_2)\\
I(\Delta_j)&=&\int_0^{\omega_c}\frac{\tanh(\frac{\sqrt{z^2+
\Delta_j^2}}{2T})}{\sqrt{z^2+\Delta_j^2}}dz.\nonumber
\end{eqnarray}
Fig.~5 shows the temperature dependences of superconducting gaps
calculated by Eq.~(1) for the values of coupling parameters
indicated in Fig.~3. Parameter $g_{11}$ defines the critical
temperature $T_c$ and larger gap $\Delta_1$, whereas $g_{22}$
affects mainly the smaller gap $\Delta_2$ and, thus, the
low-temperature behavior of the surface impedance.

Using the obtained gaps values we can calculate the temperature
dependences of conductivities
$\sigma_j(T)=\sigma'_j(T)-i\sigma''_j(T)$ in each band ($j=1,2$)
using the following generalization of formulas~\cite{Nam}:
{\small\begin{eqnarray}
\frac{\sigma_j(T)}{\sigma_j''(0)}=-\frac{i}{2}
\left[\int_{\Delta_j-\omega}^{\Delta_j}\{\textrm{I}\}\tanh(\frac{\omega
+\omega'}{2T})d\omega'+
\int_{\Delta_j}^{\infty}\left(\{\textrm{I}\}\tanh(\frac{\omega+
\omega'}{2T})-
\{\textrm{II}\}\tanh(\frac{\omega'}{2T})\right)d\omega' \right]\\
\{\textrm{I}\}=\frac{g+1}{\varepsilon_{-}-i\gamma} +
\frac{g-1}{-\varepsilon_{+}-i\gamma}\nonumber,\qquad
\{\textrm{II}\}=\frac{g+1}{\varepsilon_{-}-i\gamma} +
\frac{g-1}{\varepsilon_{+}-i\gamma}\nonumber,\qquad
g=\frac{\omega'(\omega'+\omega) + \Delta_j^2}{\sqrt{\omega'^{2} -
\Delta_j^2} \sqrt{(\omega' + \omega)^2 - \Delta_j^2}},\nonumber\\
\varepsilon_{-}=\sqrt{(\omega+\omega')^2-\Delta_j^2} -
\sqrt{(\omega')^2-\Delta_j^2},\qquad
\varepsilon_{+}=\sqrt{(\omega+\omega')^2-\Delta_j^2} +
\sqrt{(\omega')^2-\Delta_j^2}.\nonumber
\end{eqnarray}}
Then, neglecting the interband impurity scattering, which is
expected to be much weaker than intraband scattering, we get the
total microwave conductivity as a sum of conductivities (2) in two
bands~\cite{Dolgov}. The inflection of $\sigma''(T)$ curve is
mainly affected by $g_{12}$. If $g_{12}=0$, then we have two
independent BCS gaps, each opening at its own critical
temperature. The temperature dependence of $\sigma''(T)$ will
contain a sharp inflection point corresponding to the opening of
the smaller gap $\Delta_2$. If $g_{12}>0$, but rather small, then
both superconducting gaps will open at the same transition
temperature and the inflection in $\sigma''(T)$ curve will be
smoothed. The results of such calculations are shown in Fig.~3 by
solid lines. One can easily see that the change of $g_{12}$ for
samples \#1 and \#2 is in a good agreement with the evolution of
experimental $\sigma''(T)$ curves for these samples, that is, the
dirtier sample the greater interband coupling. Note also that
gap-to-$T_c$ ratios occurred to be almost the same for both
samples, namely, $\Delta_1\approx 1.8\,T_c$ and $\Delta_2\approx
T_c$.

Theoretical calculations of $\sigma'(T)$ curves with the same
parameters as in Fig.~3 and the fitting values of the relaxation
rates in two bands are shown in Fig.~4 by solid lines. The shift
of $\sigma'(T)$ maximum for sample \#2 to $T\approx 0.5\,T_c$ is
explained by the fact that the coherence peak arises from the
energy band with the smaller gap $\Delta_2$ at $\Delta_2(T)\approx
T$. In contrast to the excellent agreement between the
calculations and experimental data on $\sigma''(T)$ in Fig.~3,
only the qualitative agreement takes place for $\sigma'(T)$. The
possible reason of quantitative difference may be caused by the
simplification of the theoretical model used. For example, it does
not account for the interband scattering and strong coupling
effects.

In the absence of detailed calculations of V$_3$Si Fermi surface
we cannot treat the data in the framework of an anisotropic
$s$-wave model of superconductivity. However, the calculated
temperature dependence of NMR relaxation rate (analogue of
$\sigma'(T)$) for oblate $\Delta(\bf k)$ gap
anisotropy~\cite{Maki} does not show significant difference from
the standard BCS model. The appearance of the inflection on
$\sigma''(T)$ in the model~\cite{Maki} is also quite unlike.

In conclusion, the real and imaginary parts of the microwave
conductivity were measured in two V$_3$Si crystals with different
silicon content, purity regime and $T_c$ values. The results
obtained cannot be explained by single-band BCS model of
superconductivity. At the same time they can be well described by
two-band theory. In both samples the values of gap-to-$T_c$ ratios
are approximately the same, namely, the large gap is
$\Delta_1\approx 1.8\,T_c$ and the smaller one is $\Delta_2\approx
T_c$. This experimental evidence of two energy gaps in V$_3$Si,
which seemed earlier to be conventional BCS-superconductor, can be
confirmed also by our preliminary microwave measurements of
V$_3$Si crystals with silicon content lowered down to 20\%.

We would like to thank V.A.~Marchenko for producing V$_3$Si
crystals. Helpful discussions with A.A.~Golubov, O.V.~Dolgov and
E.G.~Maksimov are gratefully acknowledged. This research was
supported by RFBR grants Nos. 03-02-16812 and 04-02-17358.
Yu.~A.~N. thanks Russian Science Support Foundation.

\begin{figure}[h]
\centerline{\includegraphics[width=0.98\columnwidth,clip]{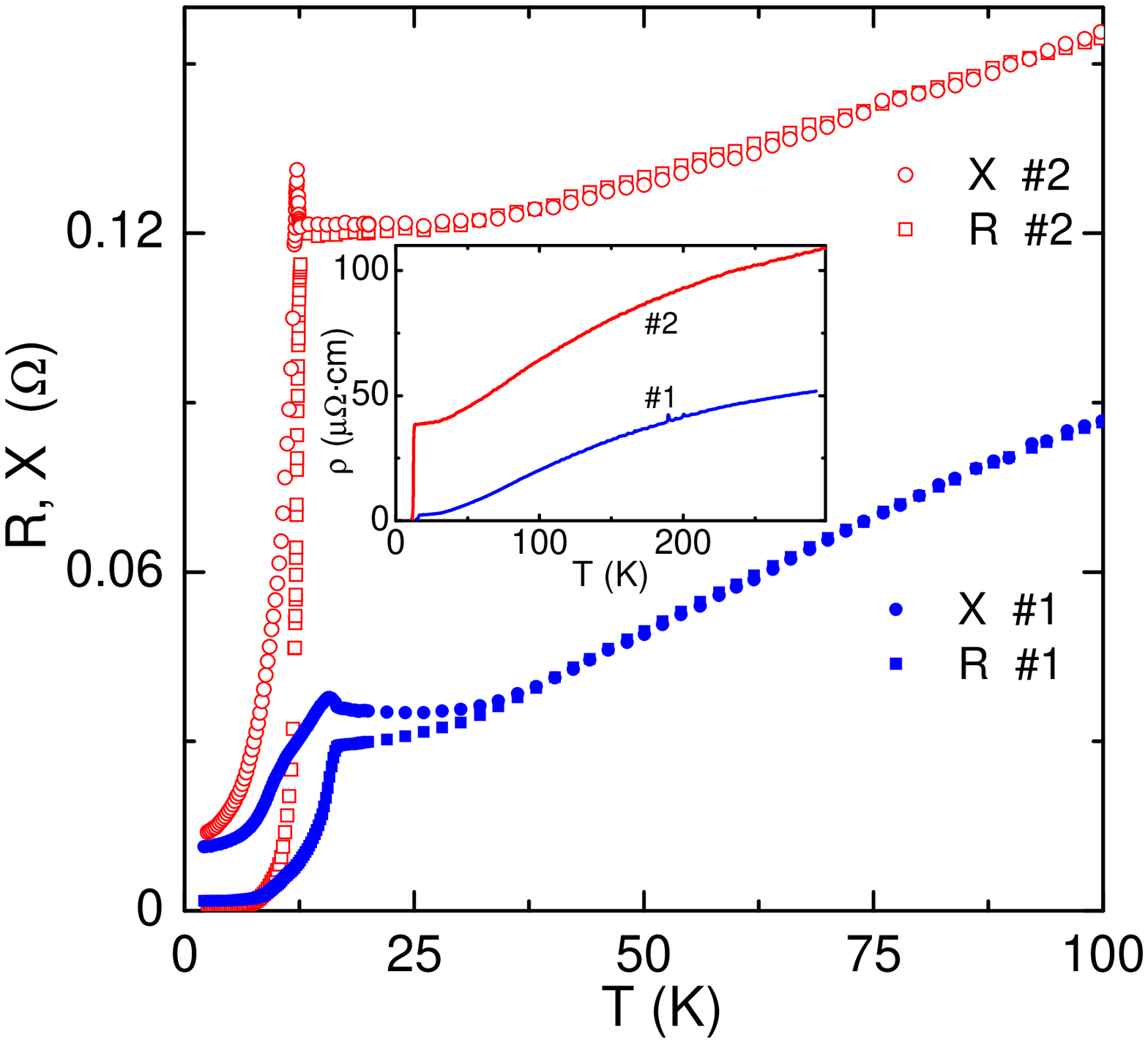}}
\caption{Temperature dependences of the surface resistance
(squares) and reactance (circles) of sample \#1 (solid symbols)
and \#2 (open symbols). Inset shows the resistivities in the
normal state of the samples.}
\end{figure}

\begin{figure}[h]
\centerline{\includegraphics[width=0.98\columnwidth,clip]{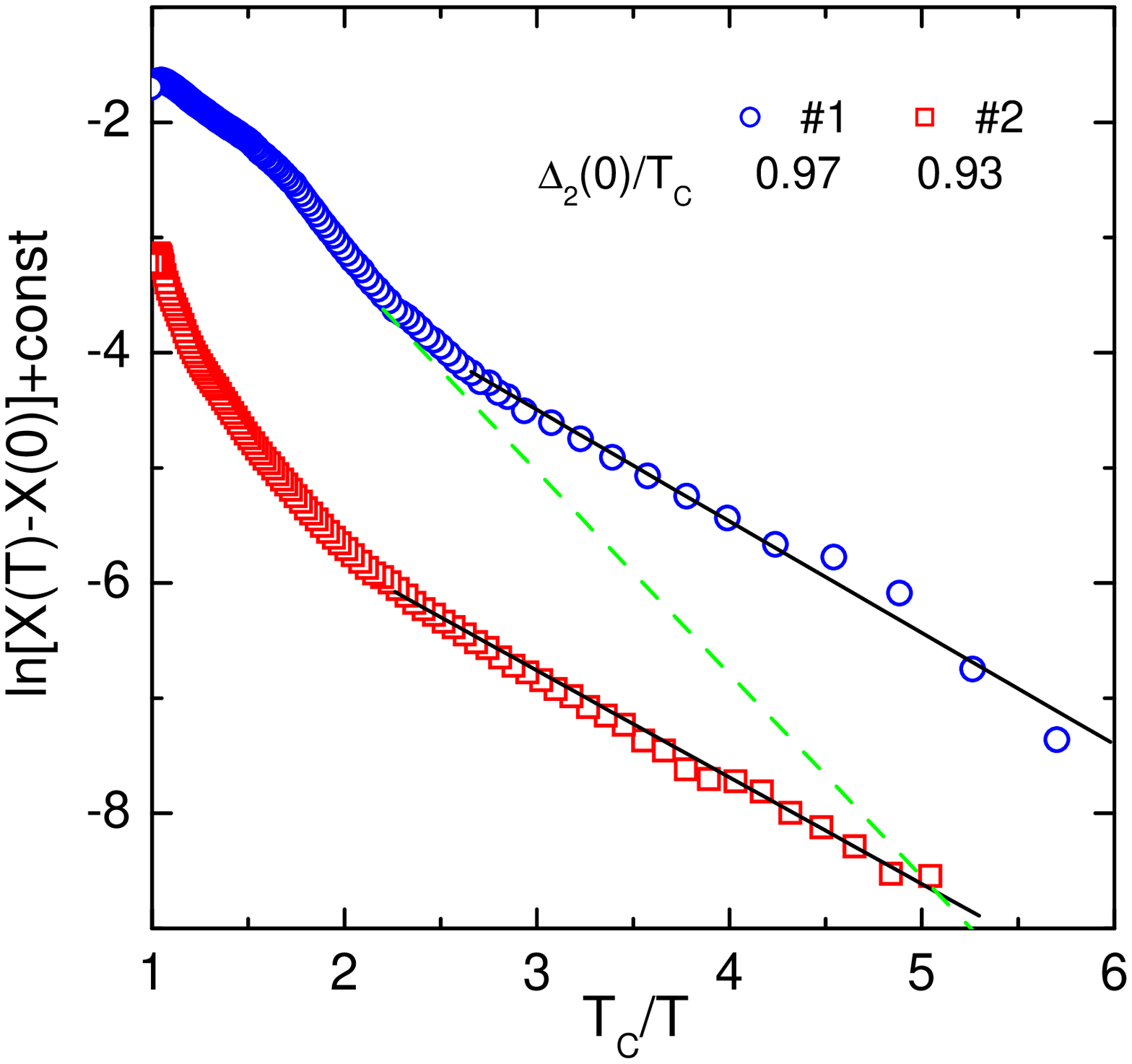}}
\caption{Surface reactance in the logarithmic scale versus $T_c/T$
(the data for sample \#1 are shifted up by 2) . The slopes of the
linear at $T<T_c/3$ sections of these curves directly give the
values of the smaller gap $\Delta_2(0)$. Single gap BCS slope is
shown by dashed line.}
\end{figure}

\begin{figure}[h]
\centerline{\includegraphics[width=0.98\columnwidth,clip]{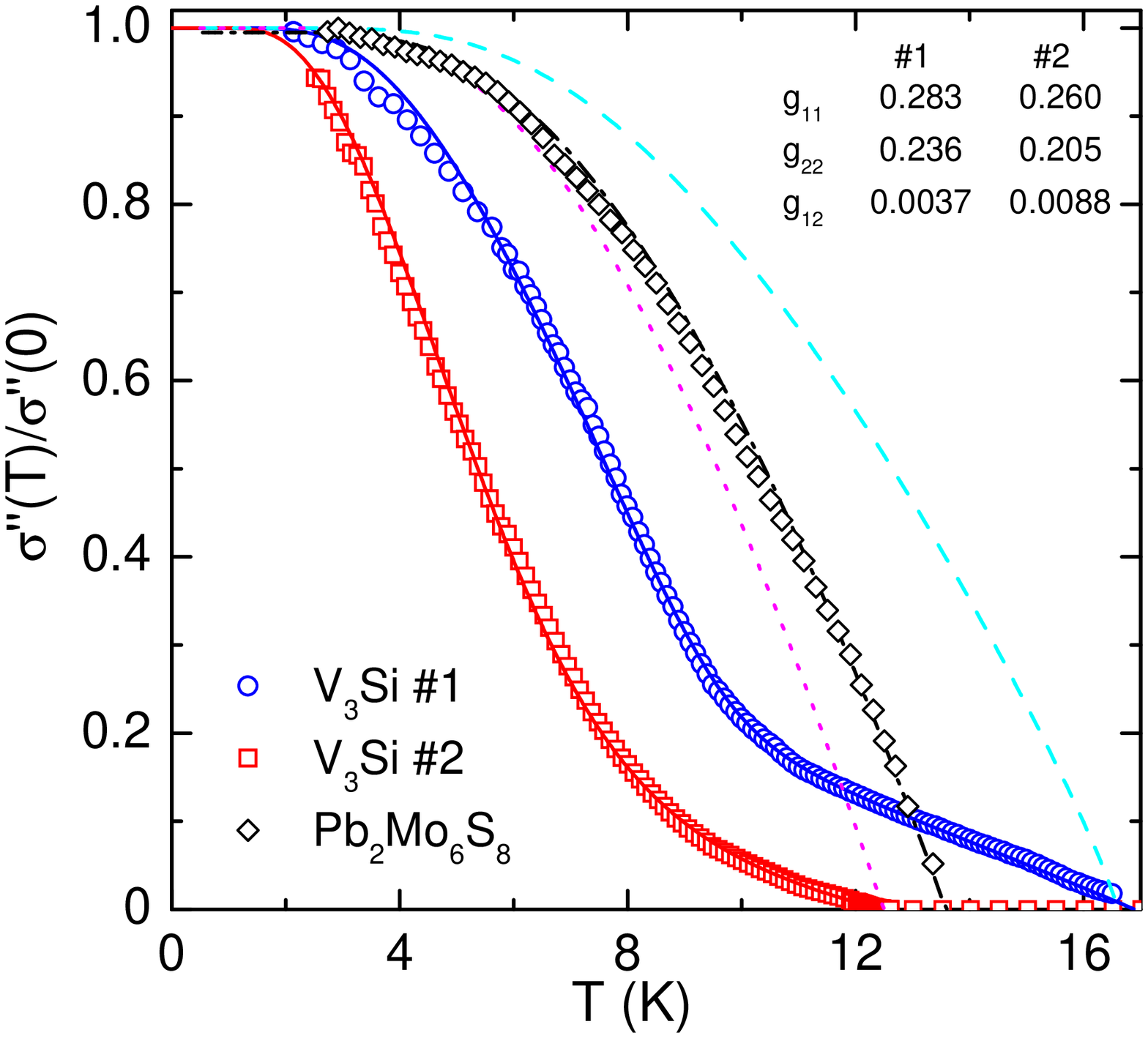}}
\caption{Experimental temperature dependences of the imaginary
part of the conductivity in samples \#1 (circles), \#2 (squares),
and Pb$_{2-x}$Mo$_6$S$_8$ crystal (diamonds). Solid lines stand
for the case of weak-coupling two-band theory with the parameters
listed in the corner. Dashed, dotted and dash-dotted lines show
single-band BCS calculations for $T_c=16.5$, 12.5, and 13.6~K
respectively.}
\end{figure}

\begin{figure}[h]
\centerline{\includegraphics[width=0.98\columnwidth,clip]{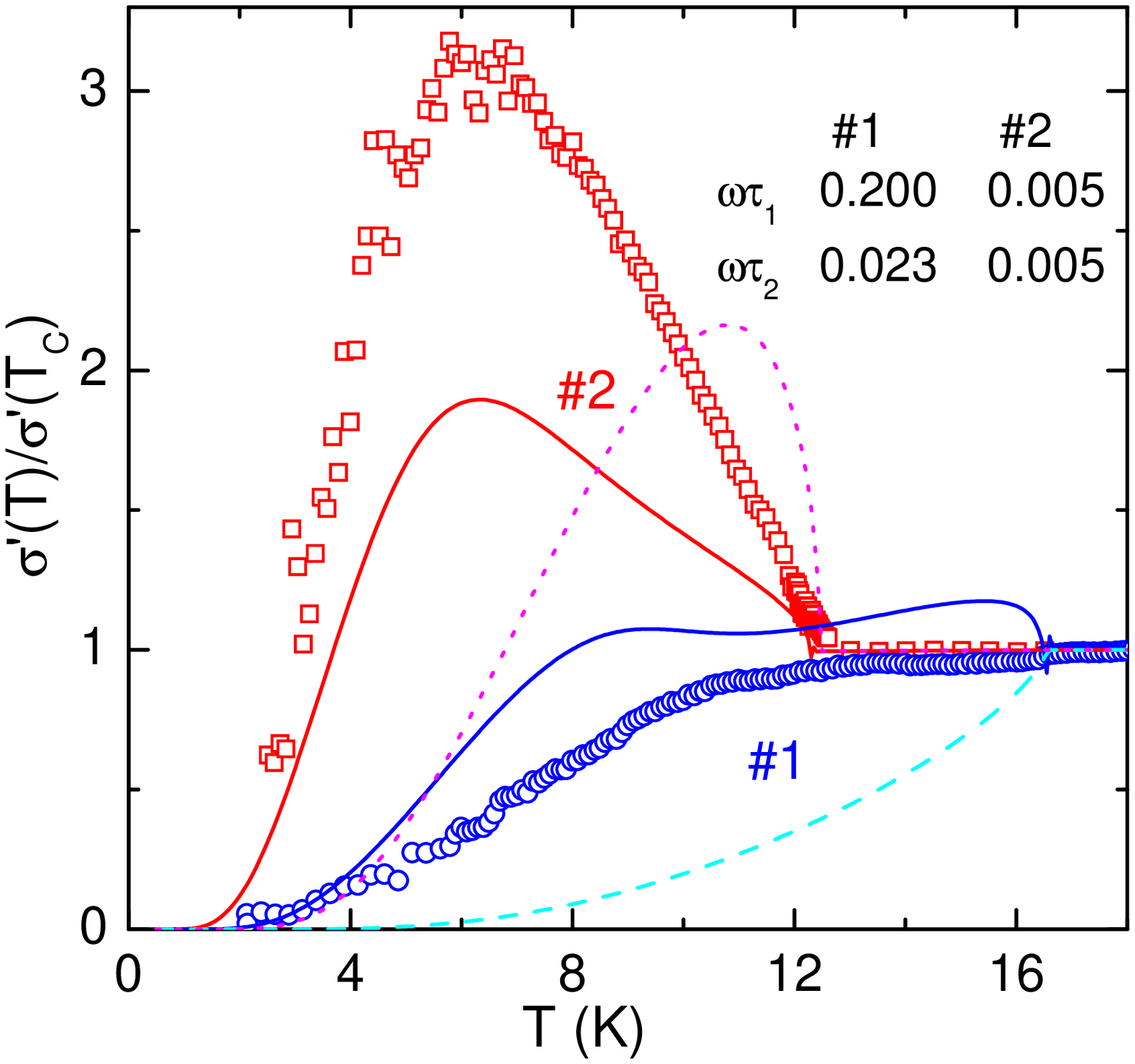}}
\caption{Temperature dependences of the real part of the
conductivity of samples \#1 (circles) and \#2 (squares). Solid
lines show the calculations in the framework of two-band model
with the same parameters as in Fig.~3. Relaxation times used for
fitting are shown in the corner of the plot. Dashed line stands
for the single-band BCS result for clean superconductor with
$T_c=16.5$~K. Dotted line corresponds to dirty BCS limit and
$T_c=12.5$~K.}
\end{figure}

\begin{figure}[h]
\centerline{\includegraphics[width=0.98\columnwidth,clip]{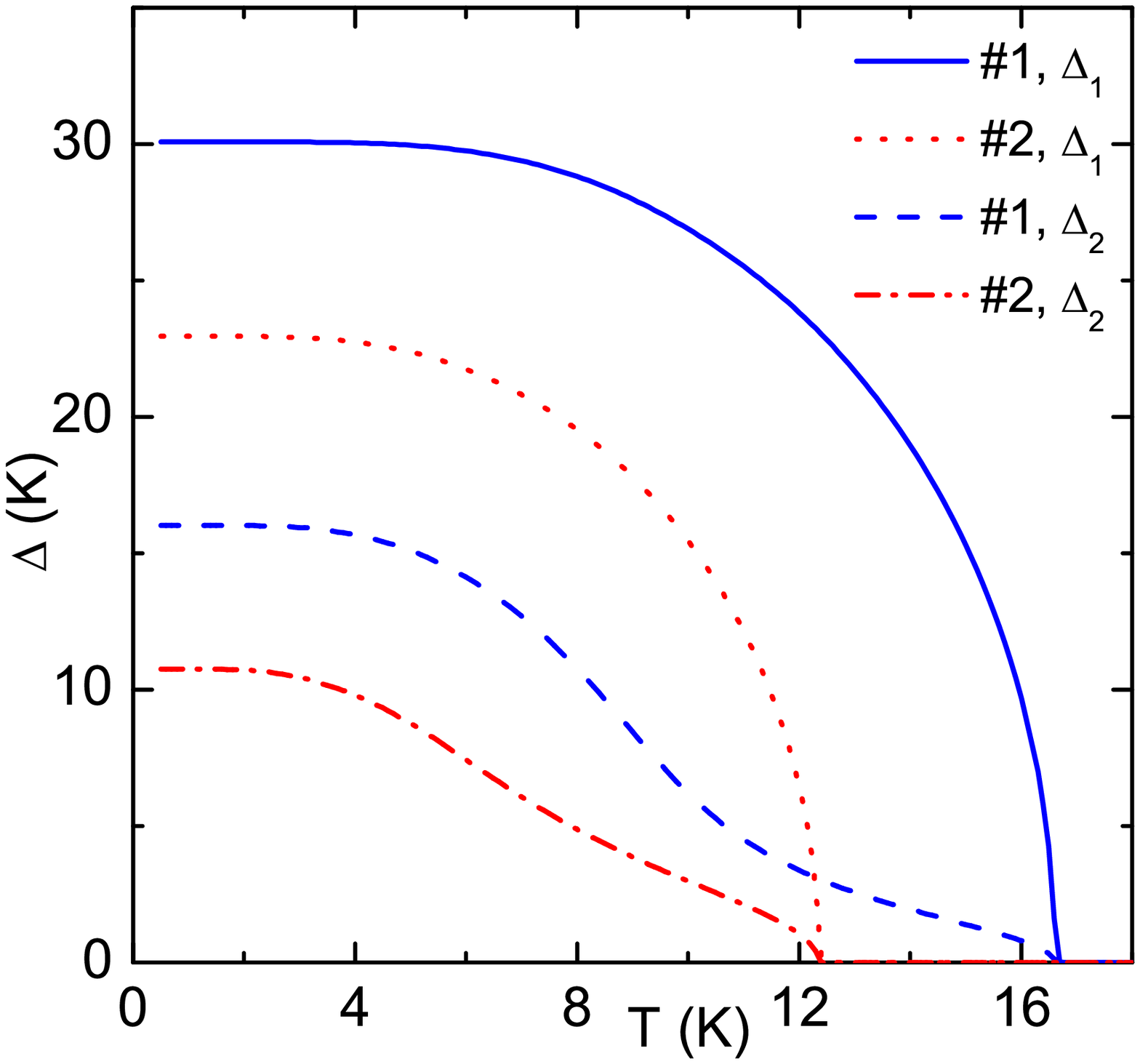}}
\caption{Solid and dashed lines show the temperature dependences
of superconducting gaps $\Delta_1$ and $\Delta_2$ calculated by
Eq.~(1) for superconductor with $T_c=16.5$~K. Dotted and
dash-dotted lines stand for the case of $T_c=12.5$~K.}
\end{figure}


\begin{thebibliography}{30}


\bibitem{BCS} J.~Bardeen, L.N.~Cooper and J.R.~Schrieffer, Phys.
Rev. {\bf 108}, 1175 (1957).

\bibitem{Mosk} V.A.~Moskalenko, Fiz. Met. Metalloved. {\bf 8}, No. 4, 503 (1959)
[Phys. Met. Metallogr. (USSR) {\bf 8}, No. 4, 25 (1959)];
V.A.~Moskalenko, M.E.~Palistrant,  Zh. Eksperim. i Teor. Fiz. {\bf
49}, 770 (1965) [Soviet Phys. JETP {\bf 22}, 536 (1966)].

\bibitem{Suhl} H.~Suhl, B.T.~Matthias and L.R.~Walker, Phys. Rev. Lett. {\bf 3}, 552 (1959).

\bibitem{Papa} B.M.~Klein, L.L.~Boyer and
D.A.~Papaconstantopoulos, L.F.~Mattheiss, Phys. Rev. B {\bf 18},
6411 (1978).

\bibitem{Bouquet} V.~Guritanu, W.~Goldacker, F.~Bouquet, Y.~Wang,
R.~Lortz, G.~Goll and A.~Junod, Phys. Rev. B {\bf 70}, 184526
(2004).

\bibitem{Boak} E.~Boaknin, M.A.~Tanatar, J.~Paglione, D.~Hawthorn,
F.~Ronning, R.W.~Hill, M.~Sutherland, L.~Taillefer, J.~Sonier,
S.M.~Hayden, and J.W.~Brill, Phys. Rev. Lett. {\bf 90}, 117003
(2003).

\bibitem{Brock} J.C.F.~Brock, Solid State
Comm. {\bf 7}, 1789 (1969).

\bibitem{Jin} B.B.~Jin, T.~Dahm, A.I.~Gubin,
E-M.~Choi, H.J.~Kim, S-IK.~Lee, W.N.~Kang, and N.~Klein, Phys.
Rev. Lett. {\bf 91}, 127006 (2003).

\bibitem{Sri7} S.~Sridhar and W.~L.~Kennedy, Rev. Sci. Instrum.
{\bf 54}, 531 (1988).

\bibitem{Tru1}
M.R.~Trunin, J. Supercond. {\bf 11}, 381 (1998).

\bibitem{Halbr}  J.P.~Turneaure, J.~Halbritter, and H.A.~Schwettman,
J. Supercond. {\bf 4}, 341 (1991).

\bibitem{Nam} S.B.~Nam, Phys. Rev. {\bf 156} 470, 487 (1967).

\bibitem{Dolgov} I.I.~Mazin, O.K.~Andersen, O.~Jepsen,
O.V.~Dolgov, J.~Kortus, A.A.~Golubov, A.B.~Kuz'menko, and D. van
der Marel, Phys. Rev. Lett. {\bf 89}, 107002 (2002).

\bibitem{Maki} A.I.~Posazhennikova, T.~Dahm, and K.~Maki,
Europhys. Lett., {\bf 60}(1), 134 (2002).

\end{thebibliography}
\end{document}